# B-Mesons Spectroscopy in Heavy Hadron Chiral Perturbation Theory

A. Upadhyay[1], M. Batra[1]
[1]SPMS, Thapar University, Patiala-147201, INDIA
*email: alka@thapar.edu, mbatra310@gmail.com

## Introduction

The new revolution in Charmonia was contemporaneous with BaBar's discovery of a narrow meson, $D_{0s}^+$ (2317). Soon after the discovery of $D_{0s}^+$ (2317) state, it got confirmed by Focus and CLEO [1] which also noticed another narrow state $D_{1s}^+$ (2463). Both these states were confirmed later by Belle. More recently other candidates have been added to the list $D_{sJ}(2860)$ by BaBar and $D_{sJ}(2710)$ by BaBar and Belle. Since then CLEO, Belle, Fermilab & BES have predicted many new states which created great enthusiasm in the Charm sector. D0 has also observed evidence for the $B^*_{s2}$ meson at a mass of (5839.1 MeV) [2]. This value has also been confirmed by CDF with higher precision [3]. But there are other measurements of excited B mesons masses $B_{s1}$ (5829.4±0.7) and $B_{s2}^*$ [2,3] reported by CDF and D0 which differ significantly and more data are needed to get at precise masses and widths. V. M. Abazov (D0 Collaboration) [2] in 2008 presented first strong evidence for resolution of excited B mesons $B_1$ and $B_2^*$. Heavy quark physics gives unique opportunity to test the prediction of Quantum Chromo dynamics and standard model for these heavy-light charm mesons. The theory of heavy meson masses, in which the symmetries of heavy and light quarks are exploited, can be used to describe the low energy interaction among heavy mesons to a better extent. Within the framework of HQET the masses, mass splitting and decay rates of heavy mesons can be studied.

## Framework

In the present work, heavy quark effective theory is combined with the chiral perturbation theory to study the spectrum of bottom meson. From the point of view of HQET, it is natural to divide quarks into two classes by comparing their Lagrangian mass with $\Lambda_{QCD}$. In this partition $m_u$, $m_d$ and $m_s$ are considered to be light where as $m_b$, $m_c$, and $m_t$ are the heavy quarks ($m_u$, $m_d$ and $m_s \ll \Lambda_{QCD} \ll m_b$, $m_c$, and $m_t$). The light quark part of the QCD Lagrangian has a chiral symmetry which arises since the current masses of the light quarks are small compared to the intrinsic mass scale of the strong interaction. Thus in the limit $m_u$, $m_d$ and $m_s \to 0$ the QCD Lagrangian for these quarks possess $SU(3)_L \otimes SU(3)_R \otimes U(1)_V$ approximate symmetry that is spontaneously broken to vector $SU(3)_V \otimes U(1)_V$ subgroup. Associated with this spontaneous breaking of the approximate chiral symmetry are the pseudoscalars octet $\pi$, k, $\eta$. The interaction of these pseudoscalar Goldstone boson with the heavy meson at low momentum can be described by an effective chiral Lagrangian that contains the most general coupling consistent with the chiral symmetry. The Lagrangian effectively describes the strong interaction of heavy-light mesons with the pseudo scalar goldstone boson at low energies. B meson masses in the heavy quark effective theory are given in terms of a single non-perturbative parameter $\bar{\Lambda}$ and non-perturbative parameters of QCD. In general, the mass of a hadron $H_Q$ containing a heavy quark Q obey an expansion of the form

$$m_X = m_Q + \bar{\Lambda} + \frac{\Delta m^2}{2m_Q} + O(\frac{1}{m_Q})$$

where X is the hadron, either in ground state (H) or an excited state (S), $m_Q$ is the mass of the heavy quark. X=H, S whereas $\Delta m^2 = -\lambda_1 + 2[J(J+1) - \frac{3}{2}]\lambda_2$. J is the total spin of meson. The two parameters $\lambda_1$ and $\lambda_2$ are non-perturbative parameters of QCD and can be estimated in various models of QCD. A good estimation of these parameters may reduce theoretical errors and uncertainties up to significant level. Although there exists several predicted values in literature [4] for $\bar{\Lambda}$ & $\lambda_1$. In all cases the values for $\lambda_1$ lie close to the 1.0 GeV. The parameters can be fit by applying constraints through experimentally well-defined masses and estimated parameter set can also be used to test the validity of other models and their predictions. The lowest and highest bounds on the parameters set can be found by using different values from the literature [5]. $\bar{\Lambda}$ & $\lambda_1$ can't be simply measured by mass measurements on dimensional grounds. $\lambda_1$ is independent of $m_Q$ and $\lambda_2$ depends on $m_Q$ logarithmically. $\lambda_1$ & $\lambda_2$ are considered to possess same values for all states in a given spin-flavor multiplet and of the order of $\Lambda_{QCD}^2$. $\lambda_1$ describes the kinetic energy term and the magnetic interaction $\lambda_2$ describes the interaction of the heavy quark spin with the gluon field and responsible for $B^* - B$ and $D^* - D$ splitting. We here apply a suitable fitting procedure using Mathematica 7.0 to find the most suitable set of all the three parameters. The parameters are here allowed to vary within their allowed values and then some of the sets that reproduce the masses with minimum error are chosen.

## Result and Discussion

The spin-flavor symmetry leads to many interesting relations between the properties of hadrons containing a heavy quark. The most direct consequences concern the spectroscopy of such states. In the $m_Q \to \infty$ limit, the spin of the heavy quark and the total angular momentum j of the light degree of freedom are separated conserved by the strong interactions. Because of heavy quark symmetry, the dynamics is independent of the spin and mass of the heavy quark. Hadronic states can thus be classified by the quantum numbers (flavor, spin, parity) of the light degrees of freedom. The spin symmetry predicts that, for fixed j ≠ 0, there is a doublet of degenerate states with total spin J $\pm \frac{1}{2}$.

Using the values from the bottom non strange sector $m_{H_1}^{(b)} = 5279.1 \pm 0.4$ MeV and $m_{H_1^*}^{(b)} = 5325.1 \pm 0.5 MeV$, $\overline{m}_{H_1}^b$ is found out to be $5313.62 \pm 0.03$ MeV. $m_{H_3}^{(b)} = 5366.3 \pm 0.6\ MeV$ given in particle data group [7] and from the relation (5) we get the spin-averaged masses of excited B- mesons

$$\overline{m}_{S_1}^{(b)} - \overline{m}_{H_1}^{(b)} = \overline{m}_{S_1}^{(c)} - \overline{m}_{H_1}^{(c)} - 56.1 \pm 25 MeV = 5705.44 \pm 48 MeV$$

Similarly for the strange bottom and charm mesons, the relation will be

$$\overline{m}_{S_3}^{(b)} - \overline{m}_{H_3}^{(b)} = \overline{m}_{S_3}^{(c)} - \overline{m}_{H_3}^{(c)} - 56.1 \pm 25 MeV = 5686.79 \pm 27 MeV$$

Equations are solved to get the values for the masses of excited B mesons. In the charm and bottom systems, one knows experimentally [6]

$$m_{B^*} - m_B \approx 46 MeV,$$
$$m_{D^*} - m_D \approx 142 MeV,$$
$$m_{D_S^*} - m_{D_S} \approx 142 MeV,$$

These mass splitting are in fact reasonably small. To be more specific, at order $1/m_Q$ one expects hyperfine corrections to resolve the degeneracy, for instance $m_{B^*} - m_B \propto 1/m_b$,. This leads to the refined prediction $m_{B^*}^2 - m_B^2 \approx m_D^2 - m_D^2 \approx const$.

$$m_{B^*}^2 - m_B^2 \approx 0.49 GeV^2, \qquad m_D^2 - m_D^2 \approx 0.55 GeV^2$$

The spin symmetry also predicts that

$$m_{B_S^*}^2 - m_{B_S}^2 \approx m_{D_S^*}^2 - m_{D_S}^2 \approx const.$$

But this constant could in principle be different from that for non-strange mesons, since the flavor quantum numbers of the light degree of freedom are different in both cases. Experimentally, however,

$$m_{D_S^*}^2 - m_{D_S}^2 \approx m_{D^*}^2 - m_D^2$$

Indicating that to first approximation hyperfine corrections are independent of the flavor of the "brown muck. One then expects the corresponding states in the bottom sector is

$$m_{B_2^*}^2 - m_{B_1}^2 \approx m_{D_2^*}^2 - m_{D_1}^2 \approx 0.17 GeV^2$$

The fact that above mass splitting is smaller for the ground-state mesons is not unexpected. For instance, in the non-relativistic constituent quark model [8] the light antiquark in these excited mesons is in a p-wave state and its wave function at the location of the heavy quark vanishes. Hence, in this model hyperfine corrections are strongly suppressed. A typical prediction of the flavor symmetry is that the "excitation energies" for states with different quantum numbers of the light degrees of freedom are approximately the same in the charm and bottom systems. For instance, one expects following splitting.

$$m_{B_S} - m_B \approx m_{D_S} - m_D \approx 100 MeV,$$
$$m_{B_1} - m_B \approx m_{D_1} - m_D \approx 557 MeV,$$
$$m_{B_2^*} - m_B \approx m_{D_2^*} - m_D \approx 593 MeV,$$

The first relation has been confirmed very nicely by the discovery of the $B_S$ meson by the ALEPH collaboration at LEP [7]. The observed mass, $m_{B_S} = 5.369 \pm 0.006 GeV$, corresponds to an excitation energy $m_{B_S} - m_B = 90 \pm 6 MeV$.

It is found that the predictions are in well agreement with the experiments. Hence if we have reliable data on the any of the parameters, quoted in the mass relations, from other experiments or theoretical models, we would be able fix the remaining parameters in the theory.

**Acknowledgments**
Part of the work is done under UGC MajorResearch project NO.SR/ 41-959/2012.

**References**
[1]. B. Aubert et al. [BaBar Collaboration], Phys. Rev. Lett. 90, 242001 (2003); E.W.Vaandering (Focus Collaboration) (2004), hep-ex/0406044; D. Besson et al. (CLEO Collaboration), AIP Conf. Proc. 698,497(2004).
[2]. CDF collaboration, CDF-Note 7938 (2005); D0 collaboration, D0-Note 5027-CONF.
[3]. V. M. Abazov [D0 Collaboration], Phys. Rev. Lett. 100, 082002(2008).
[4]. V.M. Abazov [D0 Collaboration], Phys. Rev. Lett. 99, 172001(2007); Mark B. Wise, Phys. Rev. D45, 2188-2191 (1992);
[5]. E. Jenkins, Nucl. Phys. B 412, 181(1994).
[6]. K. Nakamura et al. (Particle Data Group), J. Phys. G 37, 075021 (2010).
[7]. D. Buskulic et al. [ALEPH Collaboration], Phys. Lett. B 311,425-436, (1993).
[8]. D. Ebert, V.O. Galkin, Phys. Rev.D 57, 9(1998).